\documentclass[%
 reprint,
 amsmath,amssymb,
 aps,
floatfix,
]{revtex4-2}

\usepackage{graphicx}% Include figure files
\usepackage{dcolumn}% Align table columns on decimal point
\usepackage{bm}% bold math
\begin{document}

\preprint{APS/123-QED}

\title{Gravitational Waves Through Time: \\Scientific Significance, Detection Techniques, and Recent Breakthroughs
}% Force line breaks with \\
%\thanks{A footnote to the article title}%

\author{Chris Jia}
 \altaffiliation[Email]{}%Lines break automatically or can be forced with \\
% \author{Second Author}
 \email{chris.jia2002@utexas.edu}
\affiliation{%
 Physics and Astronomy Department at the University of Texas at Austin
}%

% \collaboration{MUSO Collaboration}%\noaffiliation

% \collaboration{CLEO Collaboration}%\noaffiliation

\date{\today}% It is always \today, today,
             %  but any date may be explicitly specified

\begin{abstract}
This article, produced as part of an undergraduate research class, aims to provide an overview of gravitational waves, though it does not offer a comprehensive review. We begin with a brief discussion regarding the history of gravitational waves, beginning even before Albert Einstein's theory of general relativity, highlighting important developments and milestones in the field. We then discuss the scientific significance of gravitational wave detections such as the verification of general relativity and key properties of black holes/neutron stars. We extend our analysis into various detection techniques including interferometer-based detectors (LIGO, Virgo, GEO600), pulsar timing arrays, and proposed space-based detectors (LISA, DECIGO, BBO). Finally, we conclude our review with a brief examination of the captivating event GW190521.

\end{abstract}

%\keywords{Suggested keywords}%Use showkeys class option if keyword
                              %display desired
\maketitle

%\tableofcontents

\section{\label{sec:level1}Introduction}
Across our endless universe, a myriad of mysteries and cosmic phenomena have captivated the imagination of humanity. Over the centuries, our understanding of the universe has evolved drastically, with one of the most recent revolutionary breakthroughs being the detection of gravitational waves\textbf{---}a phenomenon predicted by Albert Einstein's theory of general relativity over a century ago \cite{Miller2019}.

Gravitational waves (GW) are ripples in spacetime itself, caused by the disturbance of compact consolidations of energy \cite{Thorne1995}. Although only extremely massive objects such as neutron stars and black holes are able to produce gravitational waves at discernible levels for our current detectors, theoretically the acceleration of any matter can cause gravitational waves. Furthermore, gravitational waves propagate at the speed of light, adding an intriguing dimension to their nature and implications \cite{Miller2019}.

With the first detection of gravitational waves on September 14, 2015 from LIGO (Laser Interferometer Gravitational-Wave Observatory), countless achievements have been made within the fields of physics and astronomy \cite{Miller2019}. Gravitational waves tied up one of the last loose ends in the theory of general relativity at a remarkably high precision ($>$96\%) for recent detections \cite{Cervantes-Cota2016}. Our understanding of black holes and neutron stars also rely heavily on data collected from GW. Significant advancements have been achieved in understanding the properties of these massive entities, including their formation rates, angular momenta, and mass limits due to the discovery of gravitational waves \cite{Miller2019}. Furthermore, gravitational waves may provide scientists with an avenue of connecting quantum mechanics with general relativity through a hypothetical elementary particle known as gravitons \cite{Sathyaprakash2009}. Although this review will not dive into the fascinating theory of gravitons, we suggest the following articles for interested readers \cite{DYSON2013}\cite{Rovelli2001}.

In Section \ref{sec:level2} of this paper, we provide a brief history of gravitational waves, beginning even before Einstein's theory of general relativity. Subsequently, in Section \ref{sec:level3}, we dive into the scientific implications and significance of detecting gravitational waves. Section \ref{sec:level4} is dedicated to an exploration of various detectors and analytical techniques employed in the study of gravitational waves. In Section \ref{sec:level5}, we shine a spotlight on a recent and particularly captivating gravitational wave event, denoted as GW190521. 

\section{\label{sec:level2}A Brief History of Gravitational Waves}

We begin our brief history of gravitational waves in 1876, roughly 40 years before Einstein’s theory of general relativity. One of the first recorded ideas similar to the concept of GW was written by William Kingdon Clifford in his paper titled “On the Space-Theory of Matter”. In his paper he notes \cite{Clifford1878},

\begin{quote}
“(1) That small portions of space \textit{are} in fact of a nature analogous to little hills on a surface which is on average flat; namely that the ordinary laws of geometry are not valid for them. (2) That the property of being curved or distorted is continually being passed on from one portion of space to another after the manner of a wave. (3) That this variation of the curvature of space is what really happens in that phenomenon which we call the \textit{motion of matter}, whether ponderable or etherial.”
\end{quote}

The first explicit mention of gravitational waves, however, was made by Poincaré in 1905. He postulated that special relativistic gravity could have waves that proliferated at the speed of light \cite{Chen2017}. 

Finally in 1915, Einstein’s proposed his theory of general relativity. Seen as a continuation of his earlier theory of special relativity, this new theory would make waves within the physics community. Fundamentally, Einstein’s theory of general relativity asserted that gravity was not a force, as interpreted under Newton’s Law, but rather a curvature of spacetime itself \cite{Cervantes-Cota2016}.

After finalizing his theory, Einstein theorized that the existence of gravitational waves might be possible\textbf{---}perhaps in ways similar to electromagnetic waves. One way of forming electromagnetic waves is by rotating an electric dipole (formed by a positive and a negative charge separated at a distance). This oscillation would, in turn, form electromagnetic waves. For gravity, however, there is no ``opposite" to a positive mass. Without a ``gravitational dipole", Einstein ultimately pursued other avenues of possibility that could produce gravitational waves \cite{Cervantes-Cota2016}. 

After many years of deliberation, Einstein ultimately reached a conclusion in his 1936 paper co-authored alongside Nathan Rosen titled ``Do gravity waves exist”. The paper concluded with a  startling statement to many scientists at the time: \textit{gravitational plane waves cannot exist} \cite{Chen2017}. Fortunately, the story doesn't end there (or else we would not be discussing it now). 

While the aforementioned paper was waiting to be published, Einstein was made aware of some issues with the resulted obtained by the paper from his new assistant, Leopold Infeld. After some modification, he arrived at a conclusion almost opposite of his original findings. The title was revised to ``On Gravitational Waves" and provided the world one of the first mathematical solutions for gravitational waves \cite{Cervantes-Cota2016}\cite{Chen2017}. In the abstract, Einstein states \cite{EINSTEIN1937}, 

\begin{quote}
``We investigate rigorously the case of cylindrical gravitational waves. It turns out that rigorous solutions exist and that the problem reduces to the usual cylindrical waves in euclidean space."
\end{quote}

Now that gravitational waves had been theorized and a mathematical solution presented, experiments were needed to confirm this fascinating phenomenon. In 1957, the Chapel Hill conference began the race to detect gravitational waves. Over the course of the six-day conference, a motley of topics were discussed, including cosmology, radio astronomy, general relativity, and, notably, the elusive gravitational waves \cite{Cervantes-Cota2016}.

\begin{figure}
    \centering
    \includegraphics[width=8 cm]{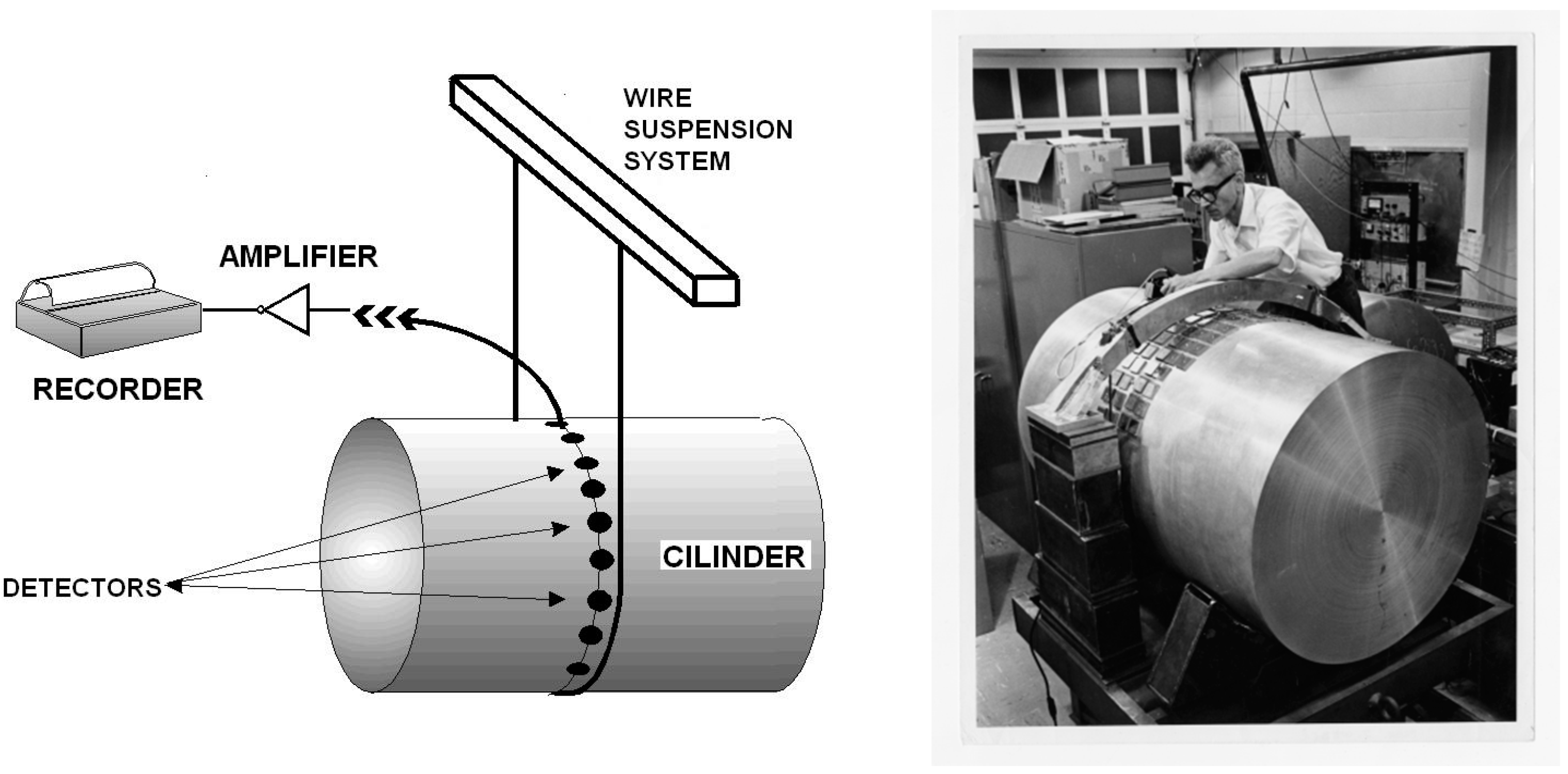}
    \caption{Image from Cervantes-Cota 2016 \cite{Cervantes-Cota2016} depicting a sketch of Weber’s cylindrical detector and a photo of Joseph Weber at the antenna.}
    \label{fig:Antenna.png}
\end{figure}

Shortly after the Chapel Hill conference, Joseph Weber began experimenting with ideas and designs for a gravitational wave detector. In 1966, he and his team built an ``antenna" surrounded by a belt of detectors. The ``antenna" was a massive aluminum cylinder and weighed roughly 3000 kg (Figure \ref{fig:Antenna.png}). To ensure the issue of ``local disturbances" such as storms or power supply fluctuations wouldn't cause false signals for the detector, he built two detectors 950 km apart and connected them with a high-speed phone line. This would ensure that the collected data could be relatively trusted and not just a result of outside factors \cite{Cervantes-Cota2016}.

\begin{figure}
    \centering
    \includegraphics[width=8.5 cm]{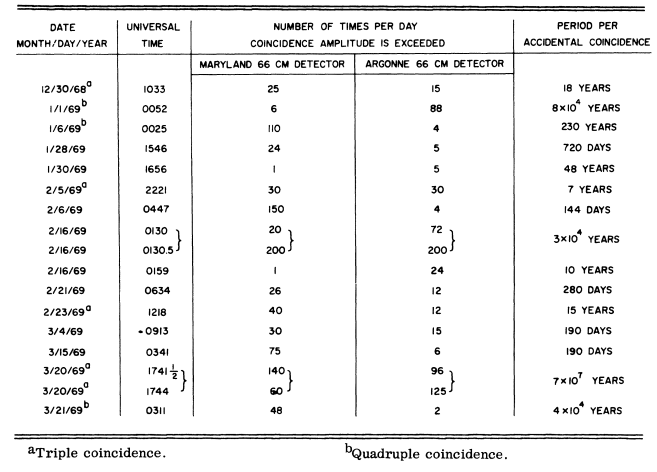}
    \caption{Table from Weber 1969 \cite{Weber1969} listing some of the ``coincidences" along with the period of time which must elapse for a coincidence with equal to or greater amplitudes to occur accidentally.}
    \label{fig:Coincidences.png}
\end{figure}

In 1969, Weber published a paper claiming he found evidence of gravitational waves. Figure \ref{fig:Coincidences.png} displays a table of several ``coincidences" between detectors. He concluded \cite{Weber1969},

\begin{quote}
``The probability that all of these coincidences were accidental is incredibly small. Experiments imply that electromagnetic and seismic effects can be ruled out with a high level of confidence. These data are consistent with the conclusion that the detectors are being excited by gravitational radiation."
\end{quote}

After much debate within the scientific community, many believed that Weber's discoveries were unreliable. His frequency of observations seemed unlikely and also appeared to be connected to extremely irregular events \cite{Cervantes-Cota2016}. 

By the mid-1970s, many new GW detectors were constructed across the world. In stark contrast to Weber's previous discoveries, these new GW detectors all arrived at a similar conclusion: \textit{there were no gravitational wave signals.} The lack of signals, however, did not deter researchers but rather invigorated them even further to try and locate gravitational waves \cite{Cervantes-Cota2016}.

At the dawn of the 21st century, several state-of-the-art gravitational wave detectors, now integral to our scientific landscape, reached completion. For interferometer-based detectors, notable names include LIGO in the USA, Virgo in Italy, and GEO600 in Germany \cite{Cervantes-Cota2016}. Other types of detectors such as pulsar timing arrays and space-based detectors have also become more prevalent in recent years and provide yet another avenue for gravitational wave detection. In Section \ref{sec:level4}, we provide a comprehensive exploration of each individual detector. Before this, however, we must focus on  the scientific significance inherent in the study of gravitational wave detections.

\section{\label{sec:level3}Scientific Significance of Gravitational Wave Detections}

This section strives to elucidate the diverse spectrum of discoveries that GW have ushered into our scientific understanding. From confirming the predictions of Einstein's theory of general relativity to understanding black holes and neutron star mergers, the exploration of gravitational wave detections opens unprecedented avenues for humanity’s understanding of the cosmos. 

\subsection{\label{sec:level3-1}Verification of general relativity}

Given Albert Einstein's pivotal role as the ``father" of gravitational waves, it is only fitting that we delve into the intricacies of how these waves serve as a validation for the principles laid out in general relativity.

Importantly, tests of general relativity (such as solar system experiments and cosmological measurements) have primarily been in low-velocity or weak-field regimes \cite{Abbott2016}. Gravitational waves, on the other hand, allow researchers to test general relativity in the high-speed and strong-field regime, an avenue previously unexplored \cite{Krishnendu2021}.

In 2020, the LIGO and Virgo collaborations examined 8 gravitational wave events within the past few years. Across 4 different tests (residuals tests, IMR consistency, parameterized tests, and modified dispersion relations), they ultimately concluded \cite{Abbott2021},

\begin{quote}
    ``We find no evidence for new physics beyond general relativity, for black hole mimickers, or for any unaccounted systematics."
\end{quote}

While a comprehensive exploration of each test lies beyond the scope of this paper, we will provide a brief overview of two specific tests: residuals tests and parameterized tests. To view the full results and efforts taken to verify general relativity using gravitational waves, we recommend the following references \cite{Krishnendu2021}\cite{Abbott2021}.

\subsubsection{\label{sec:level3-1-1} Residuals Test}

A residuals test involves examining the differences between the observed data and the model predictions (in this case, general relativity). If we assume general relativity to be correct, subtracting the astrophysical signal should reveal residuals that are comparable with instrumental noise \cite{Krishnendu2021}.

When analyzing these residuals, however, our models of general relativity are not completely perfect (although they are very accurate). Furthermore, we also don't have a complete certainty about signals in the data. Both of these factors taken together requires scientists to calculate something called the ``fitting factor" (FF) that must have a lower bound because of the uncertainties involved \cite{Krishnendu2021}.

\begin{eqnarray}
FF = \frac{||h||}{\sqrt{||h||^2 + ||s||^2}}
\label{eq:Fitting Factor 1}
\end{eqnarray}

\begin{eqnarray}
FF_{90} = \frac{SNR_{GR}}{\sqrt{SNR_{GR}^2 + SNR_{90}^2}}
\label{eq:Fitting Factor 2}
\end{eqnarray}

Equations (\ref{eq:Fitting Factor 1}) and (\ref{eq:Fitting Factor 2}) are given by \cite{Krishnendu2021}. Equation (\ref{eq:Fitting Factor 1}) represents a generic form to calculate the fitting factor with \textit{h} as the model and \textit{s} as the signal. Equation (\ref{eq:Fitting Factor 2}) is a more specific form of Equation (\ref{eq:Fitting Factor 1}) with ``90" referring to the 90\% chance that the Signal-to-Noise Ratio (SNR) of the leftover signal (residual) after removing the best-fit GR template is less than or equal to a specific value denoted as SNR$_{90}$.

\begin{figure}
    \centering
    \includegraphics[width=8.5 cm]{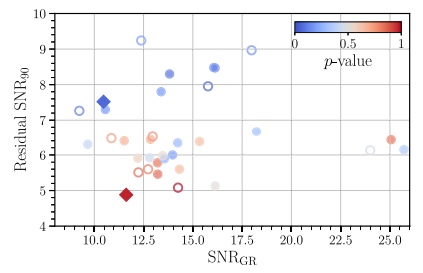}
    \caption{Plot from Abbott et al. 2021 \cite{Abbott2021} showing the SNR$_{90}$ for each GW event, as a function of the SNR maximum-likelihood template (SNR$_{GR}$), with the corresponding p-values.}
    \label{fig:Residuals_Plot.png}
\end{figure}

In Figure \ref{fig:Residuals_Plot.png}, the set of p-values show no distinct correlation between SNR$_{90}$ and SNR$_{GR}$. This implies that the SNR$_{90}$ magnitude is solely influenced by the instrumental properties associated with each event and \textbf{not} influenced by the subtracted template (general relativity). This lack of correlation provides remarkably clear evidence in favor of general relativity \cite{Abbott2021}.

\subsubsection{\label{sec:level3-1-2} Parameterized Tests}

Parameterized tests provide another avenue of verifying general relativity. In essence, these tests entail the systematic adjustment of parameters linked to the detected signals. The objective is to ascertain whether the observed gravitational wave events align with the predictions stipulated in the framework of general relativity \cite{Abbott2021}.

In the case of gravitational wave detections, scientists employ parameterized tests to evaluate various aspects of the detected signals, such as their waveform, amplitude, and frequency. These tests often involve comparing the observed data with the predictions derived from general relativity. By systematically varying specific parameters in the theoretical models and comparing the results with the actual observations, researchers can assess the consistency of the detected signals with the predictions of general relativity \cite{Abbott2021}.

The detailed investigation carried out by Abbott et al. not only reaffirms the validity of general relativity but also contributes valuable insights into the precision and accuracy of current gravitational wave detection techniques \cite{Abbott2021}. 

\subsection{\label{sec:level3-2}Insights into the nature of black holes and neutron stars}

Some of the most major events regarding gravitational waves have been mergers of black holes or binary systems of neutrons stars. The following brief sections cover the study of black holes and neutron stars from a gravitational wave perspective.

\subsubsection{\label{sec:level3-2-1} Black Holes}

The recent discoveries of major black hole merger events at LIGO (such as GW150914 and GW190521) have ignited renewed interest into the study of primordial black holes (PBHs). One of the major areas of interest is how black hole binary systems actually come about \cite{Sasaki2018}.

Two main ideas have been explored for how these pairs of black holes might form. The first involves chance encounters between PBHs within a dense environment, which works in a more recent universe. The other idea involves the tidal perturbation of distant PBHs, which could have happened in the early universe during the radiation-dominated epoch \cite{Sasaki2018}.

These investigations demonstrate the immense usefulness of gravitational waves, as it provides researchers with another avenue besides electromagnetic waves for examining the characteristics of PBHs, including their mass and abundance. As ongoing experiments receive upgrades along with the construction of future detectors, instruments will have greatly enhanced sensitivity across various frequencies. As such, the collection of more comprehensive information, including details such as eccentricity, redshift, and spatial inhomogeneities, will become readily accessible to uncover the true nature of PBHs and black hole binary systems \cite{Sasaki2018}.

\begin{figure}
    \centering
    \includegraphics[width=8.5 cm]{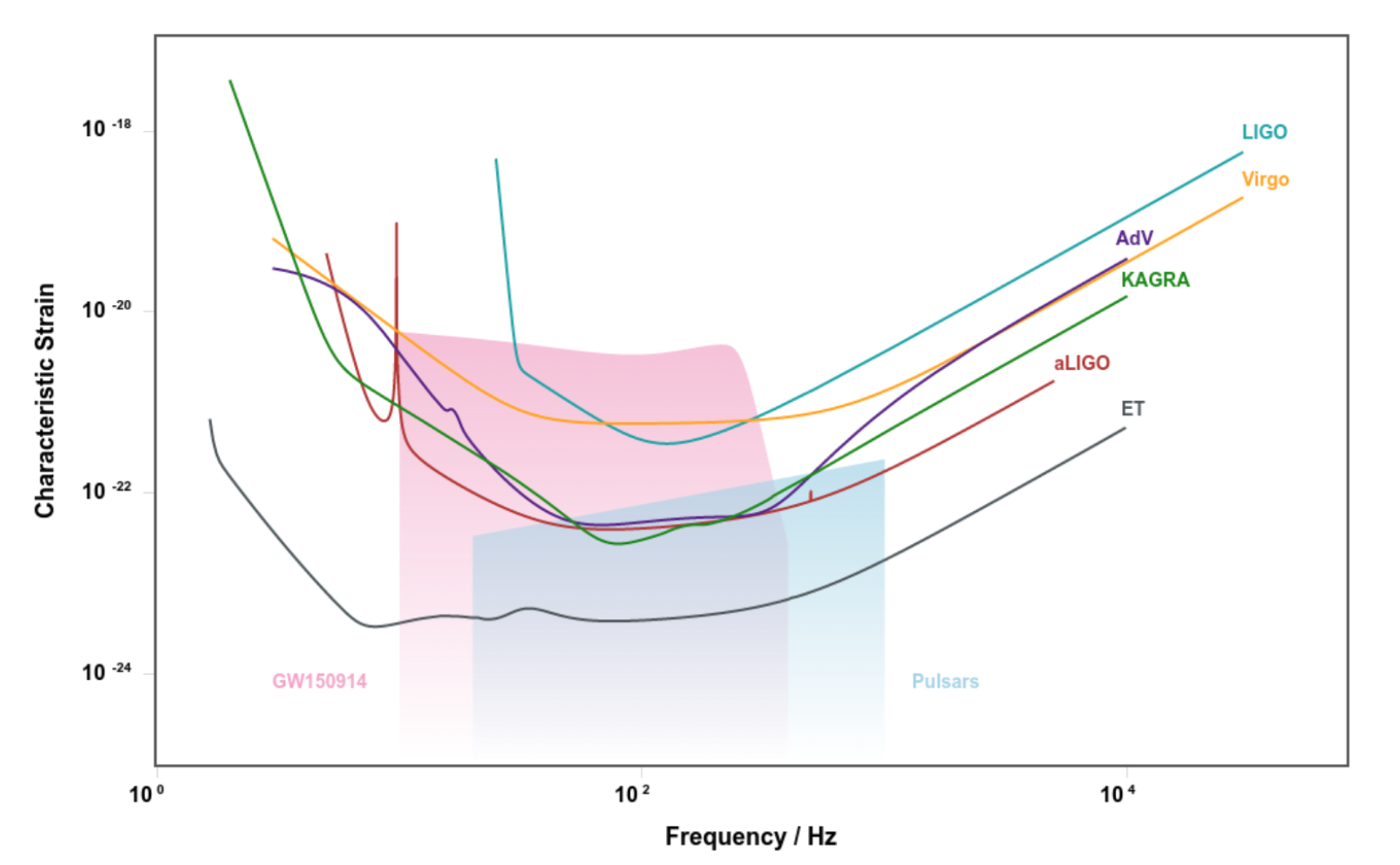}
    \caption{Figure from Sieniawska and Bejger 2019 \cite{Sieniawska2019} depicting different gravitational wave detectors and their respective range of data collection. The pink outline represents the GW150914 signal strain's amplitude ($h_0$) and frequency range, whereas the blue region represents the ranges for pulsars.}
    \label{fig:Gravitational_Wave_Amplitude.png}
\end{figure}

Figure \ref{fig:Gravitational_Wave_Amplitude.png} depicts a graph of our current and future GW detector's range of data collection in addition to the amplitude and frequency range for the first detected binary black hole coalescence (GW150914) \cite{Sieniawska2019}.

For a deeper exploration of the intricate information that gravitational waves can unveil about black holes, we highly recommend referring to the publication titled \textit{Black holes, gravitational waves and fundamental physics: a roadmap} \cite{Barack2019}.

\subsubsection{\label{sec:level3-2-2} Neutron Stars}

Just as black holes release immense gravitational waves, neutron stars also prove to be exceptional sources of these waves. Exotic processes are triggered within a neutron star's nucleus due to the tremendous pressure, which can forcefully release mass at relativistic speeds. Consequently, gravitational waves become a vital means of measuring these fascinating astrophysical objects and the dynamic processes they undergo \cite{Lasky2015}.

To generate discernible gravitational waves from rotating neutron stars, a crucial requirement is the presence of asymmetrical deformations in the neutron star. Notably, alternative mechanisms, such as pulsar ``glitches"\textbf{---}instances where neutron stars undergo a sudden ``jump" in spin frequency\textbf{---}or magnetar flares, where a neutron star with an exceptionally high magnetic dipole experiences magnetic decay, can also induce substantial gravitational waves. \cite{Lasky2015}.

Specifically, our primary emphasis will be on the asymmetrical deformations within the neutron star. Magdalena Sieniawska and Michał Bejger in 2019 provide a set of equations to estimate a GW strain amplitude \textit{h$_0$} from such neutron stars. They reach a succinct result which is given by Equation (\ref{eq: Neutron Star}) \cite{Sieniawska2019}.

\begin{eqnarray}
    h_0 \approx 10^2\frac{G\epsilon I_3\nu^2}{c^4d}
    \label{eq: Neutron Star}
\end{eqnarray}

In Equation (\ref{eq: Neutron Star}), $\epsilon$ represents a dimensionless parameter that depends on the level of asymmetry for the neutron star. The moment of inertia along a rotational axis is \textit{I$_3$} which is dependent on the neutron star's mass \textit{M} and radius \textit{R} (\textit{I$_3$ $\approx$ MR$^2$}). $\nu$ is the neutron star’s rotational frequency and \textit{d} represents the distance to the source (in this case, the neutron star) \cite{Sieniawska2019}.

As shown in Equation (\ref{eq: Neutron Star}), by measuring the amplitude of these gravitational waves and utilizing other properties we already know about a specific neutron star, we can obtain new information or verify existing information for a wide collection of neutron stars.

For more information about other particulars concerning the emission of gravitational waves from neutron stars, such as pulsar glitches and magnetar flares, we highly recommend \cite{Lasky2015}\cite{Sieniawska2019}. 

\section{\label{sec:level4}Detection Techniques}

\subsection{\label{sec:level4-1}Interferometer-based detectors}

In Section \ref{sec:level2}, we presented the first gravitational wave detector made by Joseph Weber. This sort of detector was known as a ``bar" detector, and by the 1990s, 5 major bar detectors across the globe were operational. However, these bar detectors were only able to detect gravitational waves within a certain frequency range ($\sim$700 Hz to $\sim$900 Hz) which inspired the development of other types of detectors. By 2010, interferometer-based detectors had finally finished construction (although some were in the process of receiving upgrades), and would provide a much wider scope in detectable amplitudes and frequencies of GW compared to bar detectors \cite{Riles2013}. Figure \ref{fig:Gravitational_Wave_Amplitude.png} provides a solid illustration of such ranges.

In order to detect gravitational waves through interferometers, lasers and extremely sensitive instruments are needed to detect minuscule changes. As a GW travels through space, they distort spacetime, inducing subtle stretching and compressing along perpendicular directions. Upon interacting with the detector, the wave momentarily modifies the lengths of the perpendicular arms, leading to corresponding alterations in the path length traveled by the laser beam.

\begin{figure}
    \centering
    \includegraphics[width=8.5 cm]{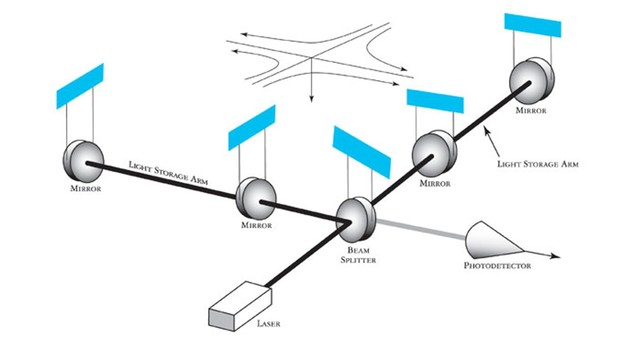}
    \caption{Diagram from Caltech/MIT/LIGO Lab \cite{Caltech} depicting a diagram of LIGO's interferometers with a gravitational waves approaching the detector}
    \label{fig:LIGO.jpg}
\end{figure}

Upon recombination, the beams may either amplify or counteract each other, leading to a noticeable shift in the interference pattern. These alterations are precisely gauged using photodetectors, which signals the presence of a gravitational wave and includes details about its characteristics such as amplitude, frequency, and arrival time \cite{Caltech}. Figure \ref{fig:LIGO.jpg} provides a simple schematic of how GW interferometers are designed.

\begin{eqnarray}
    \delta L = hL
    \label{eq:Laser Length}
\end{eqnarray}

Equation (\ref{eq:Laser Length}) shows a straightforward method of calculating the gravitational wave amplitude \textit{h}. \textit{L} refers to the length of the laser arms and correspondingly, \textit{$\delta$L} represents the resulting length difference between the two arms of the interferometer due to the gravitational wave passing through \cite{Accadia2012}.

The following sections will provide an elementary description of various interferometer-based detectors (LIGO, Virgo, and GEO600) along with major differences between them.

\subsubsection{\label{sec:level4-1-1}LIGO}

LIGO consists of two identical interferometers located in the United States: one in Livingston, Louisiana, and the other in Hanford, Washington. Each observatory features two perpendicular arms measuring 4 km in length with lasers operating at around 40 W. Power recycling mirrors, unique suspension systems, and seismic isolation techniques (passive and active) are all utilized to ensure that the detection of gravitational waves can be precisely measured. LIGO is perhaps most well-known for being the first laboratory to detect gravitational waves (GW150914) from an astrophysical source. Subsequently, the facility has continued to achieve numerous monumental detections, including GW190521, which will be further explored in Section \ref{sec:level5} \cite{Caltech}.

\subsubsection{\label{sec:level4-1-2}Virgo}

Similar to LIGO, the Virgo interferometer in Italy exhibits comparable performance with some distinct differences. Some of these differences include a shorter arm lengths (3 km versus 4 km), increased laser power (17 W versus 10 W), and seismic isolation measures. Furthermore, Virgo also excels in detecting frequencies below 40 Hz due to its seismic isolation, whereas LIGO boasts greater sensitivity around 150 Hz. Virgo's heightened sensitivity at lower frequencies also allows Virgo to detect neutron stars that LIGO can't, like Vela \cite{Riles2013}\cite{Caron1997}. Importantly, the information provided here was taken in 2013, and since then, both LIGO and Virgo have received or are in the process of receiving upgrades which aim to improve both their sensitivity and laser power. For example, Advanced Virgo can now produce laser power up to 50 W \cite{Caltech} whereas aLIGO (advanced LIGO) produces laser power up to $\sim$40 W \cite{Virgo}.

\subsubsection{\label{sec:level4-1-3}GEO600}

The GEO600 interferometer features folded arms spanning 600 meters and employs a 12 W input laser. Despite being constructed within a relatively constrained budget (and thus lacks the sensitivity that LIGO and Virgo provide), it supplies a multifaceted role within gravitational wave detection. First and foremost, it serves as a vigilant observatory during periods when LIGO and Virgo are inactive, with the potential to validate exceptionally intense event candidates. Furthermore, it also functions as a testing ground for advanced LIGO technology. By 2012, GEO600 had been primarily geared towards detecting nearby galactic supernovae \cite{Riles2013}\cite{Willke2002}. 

\subsection{\label{sec:level4-2}Pulsar timing arrays}

Pulsar timing arrays (PTAs) provide a different method for detecting low-frequency gravitational waves, as shown by Figure \ref{fig:GWspectrum.png}. As gravitational waves travel through space, they distort the fabric of spacetime just enough such that pulsar signals arrive at slightly different times on Earth. Since these pulsars provide exceptionally regular ``clocks" over an extended period of time, any minuscule deviations may hint at the presence of a gravitational wave \cite{Riles2013}.

The 3 primarily PTA collaborations include EPTA (European Pulsar Timing Array), NANOGrav (North American Nanohertz Observatory of Gravitational Waves), and PPTA (Parkes Pulsar Timing Array). Each of the telescopes hosts a PTA collaboration and conducts routine timing programs occurring at intervals ranging from 1 week to 1 month. These three collaborations collectively form the IPTA (International Pulsar Timing Array), under which they share data sets and jointly publish findings \cite{Lommen2015}.

In 2016, S. R. Taylor et al. noted in their publication that while PTAs have set upper limits on the amplitude of gravitational waves in the nanohertz range, the prospect of small pulsar groups detecting gravitational waves within the next few decades appears unlikely. Nevertheless, researchers anticipate a significantly increased likelihood of detection by employing more extensive pulsar arrays, as discussed in \cite{Taylor2016}.

Just 6 years later, the IPTA partially proved this prediction correct. They found evidence of a low-frequency gravitational wave pattern with am  amplitude and spectral index that aligned with expectations from supermassive black hole mergers. However, there was no clear evidence confirming these waves' gravitational origin \cite{Antoniadis2022}.

Continued research employing PTAs promises to enhance our understanding of gravitational waves. The collaborative efforts of international initiatives through the IPTA showcase the power of global collaborations, and the future holds great promise with ongoing advancements in technology, data-sharing practices, an expanding array of pulsars, and increased sensitivity all in hopes of detecting elusive gravitational waves \cite{Antoniadis2022}. 

\subsection{\label{sec:level4-3}Space-based detectors}

Space-based detectors represent a groundbreaking frontier in the realm of gravitational wave astronomy, offering a unique vantage point beyond the constraints of Earth's atmosphere. Unlike ground-based counterparts, space-based detectors are not hindered by terrestrial noise and environmental factors, enabling them to probe the cosmos with unprecedented precision. The following section will primarily focus on the Laser Interferometer Space Antenna (LISA) with a brief mention to other proposed space-based gravitational-wave detectors such as DECIGO and BBO. 

\subsubsection{\label{sec:level4-3-1}LISA}

\begin{figure}
    \centering
    \includegraphics[width=8.5 cm]{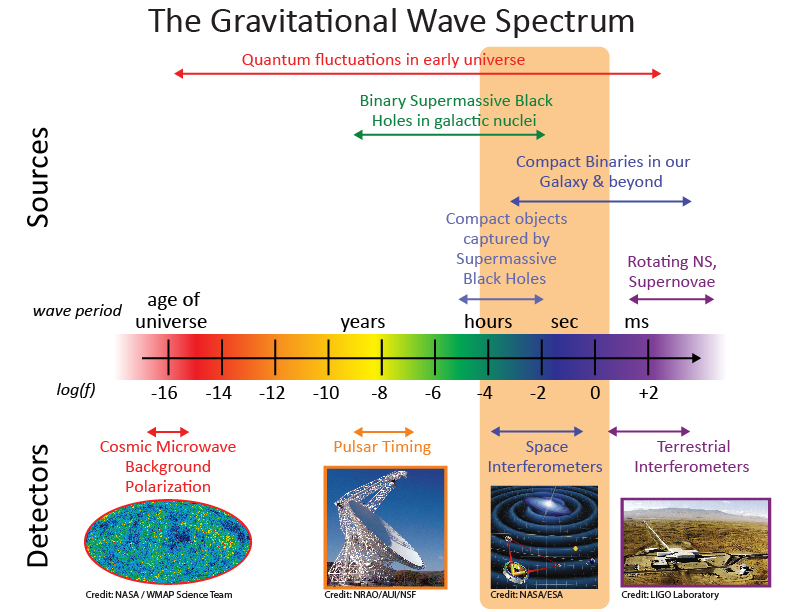}
    \caption{Image from NASA \cite{NASA} depicting the various detection techniques used to probe the gravitational wave spectrum.}
    \label{fig:GWspectrum.png}
\end{figure}

Within the expansive realm of gravitational wave frequencies, the proposed LISA mission focuses primarily on the lower frequency spectrum from around 0.1 mHz to 1 Hz (as seen in Figure \ref{fig:GWspectrum.png}). As such, LISA will be able to better detect sources with wider orbit and greater mass than LIGO, thereby expanding the range of observable gravitational wave phenomena \cite{NASA}.

\begin{figure}
    \centering
    \includegraphics[width=8.5 cm]{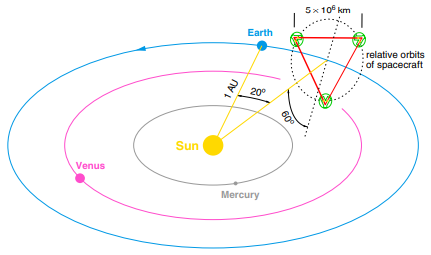}
    \caption{Illustration from Danzmann and Rudiger 2003 \cite{Danzmann2003} illustrating the proposed LISA mission. In the figure, the the triangle arms are scaled by a factor of 10.}
    \label{fig:LISA.png}
\end{figure}

LISA's planned operational configuration involves three spacecraft forming an equilateral triangle in space, with extended ``arms" (lasers) spanning about a million miles (Figure \ref{fig:LISA.png}). Because some gravitational waves have wavelengths longer than Earth, the long ``arms" of LISA would allow us to probe various new astrophysical sources and phenomenon such as ultra-compact binaries and supermassive black hole mergers. These gravitational wave events cause the spacecraft to slightly alter their respective position, which in turn produces a specific pattern from the lasers that can provide crucial insights into the gravitational wave's origin, such as the source's location and physical characteristics \cite{NASA}.

Although the LISA mission still doesn't have a definitive launch date, numerous researchers remain optimistic about the expansive capabilities that space-based detectors could potentially offer. For example, one primary objective of LISA is to identify a stochastic gravitational wave background. This background not only provides insights into our early universe right after the Big Bang, but may also reveal key details of other astrophysical sources \cite{Bartolo2022}. Additional information we may obtain from LISA include supernovae processes, galaxy formation, and the nature of dark matter/dark energy \cite{Danzmann2003}.

\subsubsection{\label{sec:level4-3-2}Other proposed space-based missions}

In addition to LISA, other notable proposed space-based detectors include DECIGO (DECI-hertz Interferometer Gravitational-wave Observatory) and BBO (Big-Bang Observer). These detectors also primarily focus in the 0.1 - 1 Hz range and would observe primordial gravitational wave background and neutron stars/black hole mergers, similar to LISA \cite{Nishizawa2012}. 

In the current preliminary design for both DECIGO and BBO, four clusters of satellites, each comprising of three spacecraft, will orbit the Sun with a sidereal year period. Similar to LISA, each spacecraft within a cluster will be connected by lasers. Two of these clusters are positioned closely together, while the other two are extensively separated along Earth's orbit. This arrangement allows for increased sensitivity to the stochastic GW background and improves our ability to identify the location of neutron star binaries. Researchers predict up to $10^5$ - $10^6$ neutron star binaries could be observed per year with these space-based detectors \cite{Nishizawa2012} \cite{Cutler2009}.

\begin{figure}
    \centering
    \includegraphics[width=8.5 cm]{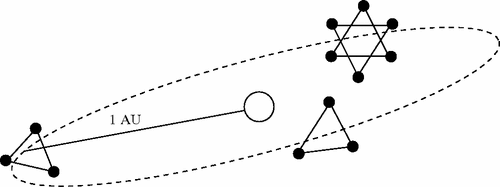}
    \caption{Illustration from Cutler and Holz 2009 \cite{Cutler2009} portraying the preliminary design for BBO. DECIGO would follow an identical composition, except the ``arms" would be 50 times shorter than BBO's.}
    \label{fig:BBO.png}
\end{figure}

Regrettably, considerable effort and funding are still required before the launch of these missions. While the potential for groundbreaking technology to yield vast amounts of new information is promising, the journey toward achieving this goal is expected to encounter several challenges along the way.

\section{\label{sec:level5}Recent Breakthrough: GW 190521}

In this last subsection, we dive into the remarkable scientific breakthrough represented by the gravitational wave event GW190521. This groundbreaking discovery, made possible by advanced gravitational wave observatories, was detected on May 21, 2019 by both LIGO and Virgo. Most notably, GW190521 was caused by the massive merger of two black holes: one at $\sim$85 solar-masses and the other at $\sim$66 solar-masses. If true, this would be the most massive black hole merger observed by our detectors to date \cite{AEI2020}.

In addition to GW190521 being an extremely massive black hole merger, it also challenges our understanding of how black holes form. Currently, there's a predicted ``mass gap" between 65 \(\textup{M}_\odot\) and 135 \(\textup{M}_\odot\) for black holes formed from dying stars \cite{DeLuca2021}. Stars within this mass range are thought to undergo pair instability at the end of their life, an unstable process in which a star leaves no remnants behind after death \cite{Abbott2020}. Although it is still possible for black holes within this mass range to  form (such as through the gradual merging of smaller black holes or theories beyond the Standard Model), GW190521 provides an interesting avenue to test such hypotheses \cite{DeLuca2021}.

Furthermore, another intriguing event may be associated with GW190521. Observations by the Zwicky Transient Facility (ZTF) regarding this event noted a flash of light, a completely unexpected result as the merger of two black holes are typically invisible to light-detecting telescopes. This suggests one of two possibilities: the black hole merger event occurred near a third supermassive black hole accretion disk or it may just be the result of another astrophysical process. If it is the first possibility, Graham et al. provides an explanation of how a burst of light could occur. After black holes merge, they ``kick" off in a random direction. If this were to happen near an accretion disk of a supermassive black hole, the newly formed black hole shoots through the cloud of gas and dust, releasing an electromagnetic flare \cite{graham2023}.

The total energy transmitted to the accretion disk $E_b$ can be described in Equation (\ref{eq:Total Energy}) from Graham et al. \cite{graham2023}.

\begin{eqnarray}
    E_b = 1/2 M_b v_k^2 = 3/2 N k_B T_b
    \label{eq:Total Energy}
\end{eqnarray}

$M_b$ = $Nm_H$ represents the mass of the bound gas (with \textit{N} defined as the number of atoms of hydrogen with mass $m_H$), $v_k$ is the ``kick" velocity of the merged black holes, $k_B$ is the Boltzmann constant, and $T_b$ signifies the average temperature of the gas ``post-shock". Furthermore, we can also estimate the luminosity associated with the accretion disk as the Bondi–Hoyle–Lyttleton (BHL) luminosity $L_{BHL}$ as seen in Equation (\ref{eq:BHL Luminosity}) where $\eta$ is the radiative efficiency \cite{graham2023}.

\begin{eqnarray}
    L_{BHL} = \eta \dot{M}_{BHL} c^2 
    \label{eq:BHL Luminosity}
\end{eqnarray}

\begin{eqnarray}
    \dot{M}_{BHL} = \frac{4\pi G^2M_{BBH}^2\rho}{v_{rel}^3}
    \label{eq:Mass BHL}
\end{eqnarray}

In Equation (\ref{eq:Mass BHL}), we calculate for the mass accretion rate $\dot{M}_{BHL}$. $M_{BBH}$ is the mass of the binary black hole merger with $\rho$ representing the gas density. $v_{rel}$ is the relative velocity which is equal to $v_k + c_s$ with $c_s$ as the gas sound speed \cite{graham2023}.

For a more technical discussion regarding GW190521, we highly recommend the following papers \cite{Abbott2020}\cite{Abbott2020_1}\cite{Gamba2023}.

\section{\label{sec:level6}Conclusion}

In conclusion, the journey through the exploration of gravitational waves has been nothing short of extraordinary. The initial notion of gravitational waves is attributed to William Kingdon Clifford in 1876, but it was Henri Poincaré in 1905 who explicitly postulated their existence. The concept gained momentum with Einstein's formulation of general relativity, but even Einstein himself was initially skeptical regarding their existence. In 1937, he revised his stance after discovering mathematical solutions for gravitational waves with help from his assistant, Leopold Infeld. The pursuit of experimental verification began in earnest in 1957, leading to Joseph Weber's pioneering efforts with a gravitational wave detector in 1966. Despite Weber's claims of gravitational waves detections in 1969, subsequent detectors worldwide in the mid-1970s found no evidence of gravitational waves, fueling a renewed commitment to their discovery. By the 21st century, advanced detectors like LIGO, Virgo, and GEO600 emerged, marking a new era in gravitational wave astronomy and culminating in the first historic detection, GW150914.

We then discuss the profound scientific significance of gravitational waves, confirming Einstein's general relativity in the relativistic and strong-field regime with complex tests, such as residual and parameterized tests. Additionally, gravitational waves provide insights into black hole and neutron star mergers, shedding light on primordial black hole formation and the dynamic processes within neutron stars. 

The landscape of gravitational wave detection techniques includes interferometer-based detectors, pulsar timing arrays, and space-based detectors, each offering unique perspectives and capabilities. Interferometer-based detectors, exemplified by LIGO, Virgo, and GEO600, utilize lasers and sensitive instruments to measure minute changes in spacetime caused by passing gravitational waves. PTAs, such as EPTA, NANOGrav, and PPTA, employ the precise timing of pulsar signals to detect low-frequency gravitational waves. LISA, a space-based detector, also aims to explore the low frequency spectrum (although not as low as PTAs), offering unique observational advantages. The LISA mission, along with proposed detectors like DECIGO and BBO, presents an opportunity to study diverse astrophysical phenomena, but many challenges and substantial efforts lie ahead in realizing the full potential of these detection methods.

The gravitational wave event GW190521, detected by LIGO and Virgo observatories in 2019, marked a significant breakthrough in astrophysics. The merger involved two black holes, one approximately 85 solar masses and the other around 66 solar masses, representing the most massive black hole merger ever recorded by our detectors. This discovery challenges existing models of black hole formation, particularly the predicted ``mass gap" between 65 and 135 solar masses. Furthermore, the unexpected observation of a flash of light associated with GW190521 by the ZTF introduces intriguing possibilities, suggesting the merger either occurred near a third supermassive black hole accretion disk or involved an unforeseen astrophysical process. This unexpected light emission, if associated with the merger, raises questions about the dynamics of black hole interactions and their potential impact on surrounding environments.

In essence, the study of gravitational waves has not only solidified our grasp of fundamental physics but has also ushered in a new era of discovery. As we continue to probe the universe with ever-improving technologies and methodologies, the future promises even more revelations and a deeper understanding of spacetime itself. 
%\nocite{*}

\bibliography{apssamp}% Produces the bibliography via BibTeX.

%apsrev4-2.bst 2019-01-14 (MD) hand-edited version of apsrev4-1.bst
%Control: key (0)
%Control: author (8) initials jnrlst
%Control: editor formatted (1) identically to author
%Control: production of article title (0) allowed
%Control: page (0) single
%Control: year (1) truncated
%Control: production of eprint (0) enabled
\providecommand{\noopsort}[1]{}\providecommand{\singleletter}[1]{#1}%
\begin{thebibliography}{37}%
\makeatletter
\providecommand \@ifxundefined [1]{%
 \@ifx{#1\undefined}
}%
\providecommand \@ifnum [1]{%
 \ifnum #1\expandafter \@firstoftwo
 \else \expandafter \@secondoftwo
 \fi
}%
\providecommand \@ifx [1]{%
 \ifx #1\expandafter \@firstoftwo
 \else \expandafter \@secondoftwo
 \fi
}%
\providecommand \natexlab [1]{#1}%
\providecommand \enquote  [1]{``#1''}%
\providecommand \bibnamefont  [1]{#1}%
\providecommand \bibfnamefont [1]{#1}%
\providecommand \citenamefont [1]{#1}%
\providecommand \href@noop [0]{\@secondoftwo}%
\providecommand \href [0]{\begingroup \@sanitize@url \@href}%
\providecommand \@href[1]{\@@startlink{#1}\@@href}%
\providecommand \@@href[1]{\endgroup#1\@@endlink}%
\providecommand \@sanitize@url [0]{\catcode `\\12\catcode `\$12\catcode `\&12\catcode `\#12\catcode `\^12\catcode `\_12\catcode `\%12\relax}%
\providecommand \@@startlink[1]{}%
\providecommand \@@endlink[0]{}%
\providecommand \url  [0]{\begingroup\@sanitize@url \@url }%
\providecommand \@url [1]{\endgroup\@href {#1}{\urlprefix }}%
\providecommand \urlprefix  [0]{URL }%
\providecommand \Eprint [0]{\href }%
\providecommand \doibase [0]{https://doi.org/}%
\providecommand \selectlanguage [0]{\@gobble}%
\providecommand \bibinfo  [0]{\@secondoftwo}%
\providecommand \bibfield  [0]{\@secondoftwo}%
\providecommand \translation [1]{[#1]}%
\providecommand \BibitemOpen [0]{}%
\providecommand \bibitemStop [0]{}%
\providecommand \bibitemNoStop [0]{.\EOS\space}%
\providecommand \EOS [0]{\spacefactor3000\relax}%
\providecommand \BibitemShut  [1]{\csname bibitem#1\endcsname}%
\let\auto@bib@innerbib\@empty
%</preamble>
\bibitem [{\citenamefont {Miller}\ and\ \citenamefont {Yunes}(2019)}]{Miller2019}%
  \BibitemOpen
  \bibfield  {author} {\bibinfo {author} {\bibfnamefont {M.~C.}\ \bibnamefont {Miller}}\ and\ \bibinfo {author} {\bibfnamefont {N.}~\bibnamefont {Yunes}},\ }\bibfield  {title} {\bibinfo {title} {The new frontier of gravitational waves},\ }\href@noop {} {\bibfield  {journal} {\bibinfo  {journal} {Nature (London)}\ }\textbf {\bibinfo {volume} {568}},\ \bibinfo {pages} {469} (\bibinfo {year} {2019})}\BibitemShut {NoStop}%
\bibitem [{\citenamefont {Thorne}(1995)}]{Thorne1995}%
  \BibitemOpen
  \bibfield  {author} {\bibinfo {author} {\bibfnamefont {K.~S.}\ \bibnamefont {Thorne}},\ }\href@noop {} {\bibinfo {title} {Gravitational waves}} (\bibinfo {year} {1995}),\ \Eprint {https://arxiv.org/abs/gr-qc/9506086} {arXiv:gr-qc/9506086 [gr-qc]} \BibitemShut {NoStop}%
\bibitem [{\citenamefont {Cervantes-Cota}\ \emph {et~al.}(2016)\citenamefont {Cervantes-Cota} \emph {et~al.}}]{Cervantes-Cota2016}%
  \BibitemOpen
  \bibfield  {author} {\bibinfo {author} {\bibfnamefont {J.~L.}\ \bibnamefont {Cervantes-Cota}} \emph {et~al.},\ }\bibfield  {title} {\bibinfo {title} {A brief history of gravitational waves},\ }\bibfield  {journal} {\bibinfo  {journal} {Universe}\ }\textbf {\bibinfo {volume} {2}},\ \href {https://doi.org/10.3390/universe2030022} {10.3390/universe2030022} (\bibinfo {year} {2016})\BibitemShut {NoStop}%
\bibitem [{\citenamefont {Sathyaprakash}\ and\ \citenamefont {Schutz}(2009)}]{Sathyaprakash2009}%
  \BibitemOpen
  \bibfield  {author} {\bibinfo {author} {\bibfnamefont {B.~S.}\ \bibnamefont {Sathyaprakash}}\ and\ \bibinfo {author} {\bibfnamefont {B.~F.}\ \bibnamefont {Schutz}},\ }\bibfield  {title} {\bibinfo {title} {Physics, astrophysics and cosmology with gravitational waves},\ }\href {https://doi.org/10.12942/lrr-2009-2} {\bibfield  {journal} {\bibinfo  {journal} {Living Rev. Relativ.}\ }\textbf {\bibinfo {volume} {12}} (\bibinfo {year} {2009})}\BibitemShut {NoStop}%
\bibitem [{\citenamefont {Dyson}(2013)}]{DYSON2013}%
  \BibitemOpen
  \bibfield  {author} {\bibinfo {author} {\bibfnamefont {F.}~\bibnamefont {Dyson}},\ }\bibfield  {title} {\bibinfo {title} {Is a graviton detectable?},\ }\href@noop {} {\bibfield  {journal} {\bibinfo  {journal} {International journal of modern physics. A, Particles and fields, gravitation, cosmology}\ }\textbf {\bibinfo {volume} {28}},\ \bibinfo {pages} {1330041} (\bibinfo {year} {2013})}\BibitemShut {NoStop}%
\bibitem [{\citenamefont {Rovelli}(2001)}]{Rovelli2001}%
  \BibitemOpen
  \bibfield  {author} {\bibinfo {author} {\bibfnamefont {C.}~\bibnamefont {Rovelli}},\ }\href@noop {} {\bibinfo {title} {Notes for a brief history of quantum gravity}} (\bibinfo {year} {2001}),\ \Eprint {https://arxiv.org/abs/gr-qc/0006061} {arXiv:gr-qc/0006061 [gr-qc]} \BibitemShut {NoStop}%
\bibitem [{\citenamefont {Clifford}(1878)}]{Clifford1878}%
  \BibitemOpen
  \bibfield  {author} {\bibinfo {author} {\bibfnamefont {W.~K.}\ \bibnamefont {Clifford}},\ }\href {https://archive.org/details/proceedingscamb06socigoog} {\bibinfo {title} {On the space-theory of matter}},\ \bibinfo {howpublished} {\url{https://archive.org/details/proceedingscamb06socigoog}} (\bibinfo {year} {1878})\BibitemShut {NoStop}%
\bibitem [{\citenamefont {Chen}\ \emph {et~al.}(2017)\citenamefont {Chen} \emph {et~al.}}]{Chen2017}%
  \BibitemOpen
  \bibfield  {author} {\bibinfo {author} {\bibfnamefont {C.}~\bibnamefont {Chen}} \emph {et~al.},\ }\bibfield  {title} {\bibinfo {title} {A brief history of gravitational wave research},\ }\href@noop {} {\bibfield  {journal} {\bibinfo  {journal} {Chinese Journal of Physics}\ }\textbf {\bibinfo {volume} {55}},\ \bibinfo {pages} {142} (\bibinfo {year} {2017})}\BibitemShut {NoStop}%
\bibitem [{\citenamefont {Einstein}\ and\ \citenamefont {Rosen}(1937)}]{EINSTEIN1937}%
  \BibitemOpen
  \bibfield  {author} {\bibinfo {author} {\bibfnamefont {A.}~\bibnamefont {Einstein}}\ and\ \bibinfo {author} {\bibfnamefont {N.}~\bibnamefont {Rosen}},\ }\bibfield  {title} {\bibinfo {title} {On gravitational waves},\ }\href {https://doi.org/https://doi.org/10.1016/S0016-0032(37)90583-0} {\bibfield  {journal} {\bibinfo  {journal} {Journal of the Franklin Institute}\ }\textbf {\bibinfo {volume} {223}},\ \bibinfo {pages} {43} (\bibinfo {year} {1937})}\BibitemShut {NoStop}%
\bibitem [{\citenamefont {Weber}(1969)}]{Weber1969}%
  \BibitemOpen
  \bibfield  {author} {\bibinfo {author} {\bibfnamefont {J.}~\bibnamefont {Weber}},\ }\bibfield  {title} {\bibinfo {title} {Evidence for discovery of gravitational radiation},\ }\href {https://doi.org/10.1103/PhysRevLett.22.1320} {\bibfield  {journal} {\bibinfo  {journal} {Phys. Rev. Lett.}\ }\textbf {\bibinfo {volume} {22}},\ \bibinfo {pages} {1320} (\bibinfo {year} {1969})}\BibitemShut {NoStop}%
\bibitem [{\citenamefont {Abbott}\ \emph {et~al.}(2016)\citenamefont {Abbott} \emph {et~al.}}]{Abbott2016}%
  \BibitemOpen
  \bibfield  {author} {\bibinfo {author} {\bibfnamefont {B.~P.}\ \bibnamefont {Abbott}} \emph {et~al.} (\bibinfo {collaboration} {LIGO Scientific and Virgo Collaborations}),\ }\bibfield  {title} {\bibinfo {title} {Tests of general relativity with gw150914},\ }\href {https://doi.org/10.1103/PhysRevLett.116.221101} {\bibfield  {journal} {\bibinfo  {journal} {Phys. Rev. Lett.}\ }\textbf {\bibinfo {volume} {116}},\ \bibinfo {pages} {221101} (\bibinfo {year} {2016})}\BibitemShut {NoStop}%
\bibitem [{\citenamefont {Krishnendu}\ and\ \citenamefont {Ohme}(2021)}]{Krishnendu2021}%
  \BibitemOpen
  \bibfield  {author} {\bibinfo {author} {\bibfnamefont {N.~V.}\ \bibnamefont {Krishnendu}}\ and\ \bibinfo {author} {\bibfnamefont {F.}~\bibnamefont {Ohme}},\ }\bibfield  {title} {\bibinfo {title} {Testing general relativity with gravitational waves: An overview},\ }\bibfield  {journal} {\bibinfo  {journal} {Universe}\ }\textbf {\bibinfo {volume} {7}},\ \href {https://doi.org/10.3390/universe7120497} {10.3390/universe7120497} (\bibinfo {year} {2021})\BibitemShut {NoStop}%
\bibitem [{\citenamefont {Abbott}\ \emph {et~al.}(2021)\citenamefont {Abbott} \emph {et~al.}}]{Abbott2021}%
  \BibitemOpen
  \bibfield  {author} {\bibinfo {author} {\bibfnamefont {R.}~\bibnamefont {Abbott}} \emph {et~al.} (\bibinfo {collaboration} {LIGO Scientific Collaboration and Virgo Collaboration}),\ }\bibfield  {title} {\bibinfo {title} {Tests of general relativity with binary black holes from the second ligo-virgo gravitational-wave transient catalog},\ }\href {https://doi.org/10.1103/PhysRevD.103.122002} {\bibfield  {journal} {\bibinfo  {journal} {Phys. Rev. D}\ }\textbf {\bibinfo {volume} {103}},\ \bibinfo {pages} {122002} (\bibinfo {year} {2021})}\BibitemShut {NoStop}%
\bibitem [{\citenamefont {Sasaki}\ \emph {et~al.}(2018)\citenamefont {Sasaki} \emph {et~al.}}]{Sasaki2018}%
  \BibitemOpen
  \bibfield  {author} {\bibinfo {author} {\bibfnamefont {M.}~\bibnamefont {Sasaki}} \emph {et~al.},\ }\bibfield  {title} {\bibinfo {title} {Primordial black holes—perspectives in gravitational wave astronomy},\ }\href@noop {} {\bibfield  {journal} {\bibinfo  {journal} {Classical and Quantum Gravity}\ }\textbf {\bibinfo {volume} {35}},\ \bibinfo {pages} {063001} (\bibinfo {year} {2018})}\BibitemShut {NoStop}%
\bibitem [{\citenamefont {Sieniawska}\ and\ \citenamefont {Bejger}(2019)}]{Sieniawska2019}%
  \BibitemOpen
  \bibfield  {author} {\bibinfo {author} {\bibfnamefont {M.}~\bibnamefont {Sieniawska}}\ and\ \bibinfo {author} {\bibfnamefont {M.}~\bibnamefont {Bejger}},\ }\bibfield  {title} {\bibinfo {title} {Continuous gravitational waves from neutron stars: Current status and prospects},\ }\bibfield  {journal} {\bibinfo  {journal} {Universe}\ }\textbf {\bibinfo {volume} {5}},\ \href {https://doi.org/10.3390/universe5110217} {10.3390/universe5110217} (\bibinfo {year} {2019})\BibitemShut {NoStop}%
\bibitem [{\citenamefont {Barack}\ \emph {et~al.}(2019)\citenamefont {Barack} \emph {et~al.}}]{Barack2019}%
  \BibitemOpen
  \bibfield  {author} {\bibinfo {author} {\bibfnamefont {L.}~\bibnamefont {Barack}} \emph {et~al.},\ }\bibfield  {title} {\bibinfo {title} {Black holes, gravitational waves and fundamental physics: a roadmap},\ }\href {https://doi.org/10.1088/1361-6382/ab0587} {\bibfield  {journal} {\bibinfo  {journal} {Classical and Quantum Gravity}\ }\textbf {\bibinfo {volume} {36}},\ \bibinfo {pages} {143001} (\bibinfo {year} {2019})}\BibitemShut {NoStop}%
\bibitem [{\citenamefont {Lasky}(2015)}]{Lasky2015}%
  \BibitemOpen
  \bibfield  {author} {\bibinfo {author} {\bibfnamefont {P.~D.}\ \bibnamefont {Lasky}},\ }\bibfield  {title} {\bibinfo {title} {Gravitational waves from neutron stars: A review},\ }\href {https://doi.org/10.1017/pasa.2015.35} {\bibfield  {journal} {\bibinfo  {journal} {Publications of the Astronomical Society of Australia}\ }\textbf {\bibinfo {volume} {32}},\ \bibinfo {pages} {e034} (\bibinfo {year} {2015})}\BibitemShut {NoStop}%
\bibitem [{\citenamefont {Riles}(2013)}]{Riles2013}%
  \BibitemOpen
  \bibfield  {author} {\bibinfo {author} {\bibfnamefont {K.}~\bibnamefont {Riles}},\ }\bibfield  {title} {\bibinfo {title} {Gravitational waves: Sources, detectors and searches},\ }\href {https://doi.org/10.1016/j.ppnp.2012.08.001} {\bibfield  {journal} {\bibinfo  {journal} {Progress in Particle and Nuclear Physics}\ }\textbf {\bibinfo {volume} {68}},\ \bibinfo {pages} {1–54} (\bibinfo {year} {2013})}\BibitemShut {NoStop}%
\bibitem [{\citenamefont {Caltech}\ and\ \citenamefont {MIT}()}]{Caltech}%
  \BibitemOpen
  \bibfield  {author} {\bibinfo {author} {\bibnamefont {Caltech}}\ and\ \bibinfo {author} {\bibnamefont {MIT}},\ }\href {https://www.ligo.caltech.edu/page/what-is-interferometer} {}\bibinfo {howpublished} {\url{https://www.ligo.caltech.edu/page/what-is-interferometer}}\BibitemShut {NoStop}%
\bibitem [{\citenamefont {Accadia}\ \emph {et~al.}(2012)\citenamefont {Accadia} \emph {et~al.}}]{Accadia2012}%
  \BibitemOpen
  \bibfield  {author} {\bibinfo {author} {\bibfnamefont {T.}~\bibnamefont {Accadia}} \emph {et~al.},\ }\bibfield  {title} {\bibinfo {title} {Virgo: a laser interferometer to detect gravitational waves},\ }\href {https://doi.org/10.1088/1748-0221/7/03/P03012} {\bibfield  {journal} {\bibinfo  {journal} {Journal {of} Instrumentation}\ }\textbf {\bibinfo {volume} {7}},\ \bibinfo {pages} {P03012} (\bibinfo {year} {2012})}\BibitemShut {NoStop}%
\bibitem [{\citenamefont {Caron}\ \emph {et~al.}(1997)\citenamefont {Caron} \emph {et~al.}}]{Caron1997}%
  \BibitemOpen
  \bibfield  {author} {\bibinfo {author} {\bibfnamefont {B.}~\bibnamefont {Caron}} \emph {et~al.},\ }\bibfield  {title} {\bibinfo {title} {The virgo interferometer for gravitational wave detection},\ }\href {https://doi.org/https://doi.org/10.1016/S0920-5632(97)00109-6} {\bibfield  {journal} {\bibinfo  {journal} {Nuclear Physics B - Proceedings Supplements}\ }\textbf {\bibinfo {volume} {54}},\ \bibinfo {pages} {167} (\bibinfo {year} {1997})}\BibitemShut {NoStop}%
\bibitem [{\citenamefont {Virgo}()}]{Virgo}%
  \BibitemOpen
  \bibfield  {author} {\bibinfo {author} {\bibnamefont {Virgo}},\ }\href {https://www.virgo-gw.eu/Laser/upgrade_1_F.html} {}\bibinfo {howpublished} {\url{https://www.virgo-gw.eu/Laser/upgrade_1_F.html}}\BibitemShut {NoStop}%
\bibitem [{\citenamefont {Willke}\ \emph {et~al.}(2002)\citenamefont {Willke} \emph {et~al.}}]{Willke2002}%
  \BibitemOpen
  \bibfield  {author} {\bibinfo {author} {\bibfnamefont {B.}~\bibnamefont {Willke}} \emph {et~al.},\ }\bibfield  {title} {\bibinfo {title} {The geo 600 gravitational wave detector},\ }\href {https://doi.org/10.1088/0264-9381/19/7/321} {\bibfield  {journal} {\bibinfo  {journal} {Classical and Quantum Gravity}\ }\textbf {\bibinfo {volume} {19}},\ \bibinfo {pages} {1377} (\bibinfo {year} {2002})}\BibitemShut {NoStop}%
\bibitem [{\citenamefont {Lommen}(2015)}]{Lommen2015}%
  \BibitemOpen
  \bibfield  {author} {\bibinfo {author} {\bibfnamefont {A.~N.}\ \bibnamefont {Lommen}},\ }\bibfield  {title} {\bibinfo {title} {Pulsar timing arrays: the promise of gravitational wave detection},\ }\href {https://doi.org/10.1088/0034-4885/78/12/124901} {\bibfield  {journal} {\bibinfo  {journal} {Reports on Progress in Physics}\ }\textbf {\bibinfo {volume} {78}},\ \bibinfo {pages} {124901} (\bibinfo {year} {2015})}\BibitemShut {NoStop}%
\bibitem [{\citenamefont {Taylor}\ \emph {et~al.}(2016)\citenamefont {Taylor}, \citenamefont {Vallisneri}, \citenamefont {Ellis}, \citenamefont {Mingarelli}, \citenamefont {Lazio},\ and\ \citenamefont {van Haasteren}}]{Taylor2016}%
  \BibitemOpen
  \bibfield  {author} {\bibinfo {author} {\bibfnamefont {S.~R.}\ \bibnamefont {Taylor}}, \bibinfo {author} {\bibfnamefont {M.}~\bibnamefont {Vallisneri}}, \bibinfo {author} {\bibfnamefont {J.~A.}\ \bibnamefont {Ellis}}, \bibinfo {author} {\bibfnamefont {C.~M.~F.}\ \bibnamefont {Mingarelli}}, \bibinfo {author} {\bibfnamefont {T.~J.~W.}\ \bibnamefont {Lazio}},\ and\ \bibinfo {author} {\bibfnamefont {R.}~\bibnamefont {van Haasteren}},\ }\bibfield  {title} {\bibinfo {title} {Are we there yet? time to detection of nanohertz gravitational waves based on pulsar-timing array limits},\ }\href {https://doi.org/10.3847/2041-8205/819/1/L6} {\bibfield  {journal} {\bibinfo  {journal} {The Astrophysical Journal Letters}\ }\textbf {\bibinfo {volume} {819}},\ \bibinfo {pages} {L6} (\bibinfo {year} {2016})}\BibitemShut {NoStop}%
\bibitem [{\citenamefont {Antoniadis}\ \emph {et~al.}(2022)\citenamefont {Antoniadis} \emph {et~al.}}]{Antoniadis2022}%
  \BibitemOpen
  \bibfield  {author} {\bibinfo {author} {\bibfnamefont {J.}~\bibnamefont {Antoniadis}} \emph {et~al.},\ }\bibfield  {title} {\bibinfo {title} {The international pulsar timing array second data release: Search for an isotropic gravitational wave background},\ }\href@noop {} {\bibfield  {journal} {\bibinfo  {journal} {Monthly notices of the Royal Astronomical Society}\ }\textbf {\bibinfo {volume} {510}},\ \bibinfo {pages} {4873} (\bibinfo {year} {2022})}\BibitemShut {NoStop}%
\bibitem [{\citenamefont {NASA}()}]{NASA}%
  \BibitemOpen
  \bibfield  {author} {\bibinfo {author} {\bibnamefont {NASA}},\ }\href {https://lisa.nasa.gov/#why} {}\bibinfo {howpublished} {\url{https://lisa.nasa.gov/#why}}\BibitemShut {NoStop}%
\bibitem [{\citenamefont {Danzmann}\ and\ \citenamefont {Rüdiger}(2003)}]{Danzmann2003}%
  \BibitemOpen
  \bibfield  {author} {\bibinfo {author} {\bibfnamefont {K.}~\bibnamefont {Danzmann}}\ and\ \bibinfo {author} {\bibfnamefont {A.}~\bibnamefont {Rüdiger}},\ }\bibfield  {title} {\bibinfo {title} {Lisa technology—concept, status, prospects},\ }\href {https://doi.org/10.1088/0264-9381/20/10/301} {\bibfield  {journal} {\bibinfo  {journal} {Classical and Quantum Gravity}\ }\textbf {\bibinfo {volume} {20}},\ \bibinfo {pages} {S1} (\bibinfo {year} {2003})}\BibitemShut {NoStop}%
\bibitem [{\citenamefont {Bartolo}\ \emph {et~al.}(2022)\citenamefont {Bartolo} \emph {et~al.}}]{Bartolo2022}%
  \BibitemOpen
  \bibfield  {author} {\bibinfo {author} {\bibfnamefont {N.}~\bibnamefont {Bartolo}} \emph {et~al.},\ }\bibfield  {title} {\bibinfo {title} {Probing anisotropies of the stochastic gravitational wave background with lisa},\ }\href@noop {} {\bibfield  {journal} {\bibinfo  {journal} {Journal of cosmology and astroparticle physics}\ }\textbf {\bibinfo {volume} {2022}},\ \bibinfo {pages} {9} (\bibinfo {year} {2022})}\BibitemShut {NoStop}%
\bibitem [{\citenamefont {Nishizawa}\ \emph {et~al.}(2012)\citenamefont {Nishizawa} \emph {et~al.}}]{Nishizawa2012}%
  \BibitemOpen
  \bibfield  {author} {\bibinfo {author} {\bibfnamefont {A.}~\bibnamefont {Nishizawa}} \emph {et~al.},\ }\bibfield  {title} {\bibinfo {title} {Cosmology with space-based gravitational-wave detectors: Dark energy and primordial gravitational waves},\ }\href@noop {} {\bibfield  {journal} {\bibinfo  {journal} {Physical review. D, Particles, fields, gravitation, and cosmology}\ }\textbf {\bibinfo {volume} {85}} (\bibinfo {year} {2012})}\BibitemShut {NoStop}%
\bibitem [{\citenamefont {Cutler}\ and\ \citenamefont {Holz}(2009)}]{Cutler2009}%
  \BibitemOpen
  \bibfield  {author} {\bibinfo {author} {\bibfnamefont {C.}~\bibnamefont {Cutler}}\ and\ \bibinfo {author} {\bibfnamefont {D.~E.}\ \bibnamefont {Holz}},\ }\bibfield  {title} {\bibinfo {title} {Ultrahigh precision cosmology from gravitational waves},\ }\href@noop {} {\bibfield  {journal} {\bibinfo  {journal} {Physical review. D, Particles, fields, gravitation, and cosmology}\ }\textbf {\bibinfo {volume} {80}} (\bibinfo {year} {2009})}\BibitemShut {NoStop}%
\bibitem [{AEI(2020)}]{AEI2020}%
  \BibitemOpen
  \href@noop {} {}\bibinfo {howpublished} {\url{https://www.aei.mpg.de/296843/ligo-and-virgo-catch-their-biggest-fish-so-far}} (\bibinfo {year} {2020})\BibitemShut {NoStop}%
\bibitem [{\citenamefont {De~Luca}\ \emph {et~al.}(2021)\citenamefont {De~Luca} \emph {et~al.}}]{DeLuca2021}%
  \BibitemOpen
  \bibfield  {author} {\bibinfo {author} {\bibfnamefont {V.}~\bibnamefont {De~Luca}} \emph {et~al.},\ }\bibfield  {title} {\bibinfo {title} {Gw190521 mass gap event and the primordial black hole scenario},\ }\href@noop {} {\bibfield  {journal} {\bibinfo  {journal} {Physical review letters}\ }\textbf {\bibinfo {volume} {126}},\ \bibinfo {pages} {051101} (\bibinfo {year} {2021})}\BibitemShut {NoStop}%
\bibitem [{\citenamefont {Abbott}\ \emph {et~al.}(2020{\natexlab{a}})\citenamefont {Abbott} \emph {et~al.}}]{Abbott2020}%
  \BibitemOpen
  \bibfield  {author} {\bibinfo {author} {\bibfnamefont {R.}~\bibnamefont {Abbott}} \emph {et~al.} (\bibinfo {collaboration} {LIGO Scientific Collaboration and Virgo Collaboration}),\ }\bibfield  {title} {\bibinfo {title} {Gw190521: A binary black hole merger with a total mass of $150\text{ }\text{ }{M}_{\ensuremath{\bigodot}}$},\ }\href {https://doi.org/10.1103/PhysRevLett.125.101102} {\bibfield  {journal} {\bibinfo  {journal} {Phys. Rev. Lett.}\ }\textbf {\bibinfo {volume} {125}},\ \bibinfo {pages} {101102} (\bibinfo {year} {2020}{\natexlab{a}})}\BibitemShut {NoStop}%
\bibitem [{\citenamefont {Graham}\ \emph {et~al.}(2023)\citenamefont {Graham} \emph {et~al.}}]{graham2023}%
  \BibitemOpen
  \bibfield  {author} {\bibinfo {author} {\bibfnamefont {M.~J.}\ \bibnamefont {Graham}} \emph {et~al.},\ }\bibfield  {title} {\bibinfo {title} {{Candidate Electromagnetic Counterpart to the Binary Black Hole Merger Gravitational-Wave Event S190521g}},\ }\bibfield  {journal} {\bibinfo  {journal} {PHYSICAL REVIEW LETTERS}\ }\href {https://doi.org/10.1103/PhysRevLett.124.251102} {10.1103/PhysRevLett.124.251102} (\bibinfo {year} {2023})\BibitemShut {NoStop}%
\bibitem [{\citenamefont {Abbott}\ \emph {et~al.}(2020{\natexlab{b}})\citenamefont {Abbott} \emph {et~al.}}]{Abbott2020_1}%
  \BibitemOpen
  \bibfield  {author} {\bibinfo {author} {\bibfnamefont {R.}~\bibnamefont {Abbott}} \emph {et~al.},\ }\bibfield  {title} {\bibinfo {title} {Properties and astrophysical implications of the 150 solar-mass binary black hole merger gw190521},\ }\href {https://doi.org/10.3847/2041-8213/aba493} {\bibfield  {journal} {\bibinfo  {journal} {The Astrophysical Journal Letters}\ }\textbf {\bibinfo {volume} {900}},\ \bibinfo {pages} {L13} (\bibinfo {year} {2020}{\natexlab{b}})}\BibitemShut {NoStop}%
\bibitem [{\citenamefont {Gamba}\ \emph {et~al.}(2023)\citenamefont {Gamba}, \citenamefont {Breschi}, \citenamefont {Carullo}, \citenamefont {Albanesi}, \citenamefont {Rettegno}, \citenamefont {Bernuzzi},\ and\ \citenamefont {Nagar}}]{Gamba2023}%
  \BibitemOpen
  \bibfield  {author} {\bibinfo {author} {\bibfnamefont {R.}~\bibnamefont {Gamba}}, \bibinfo {author} {\bibfnamefont {M.}~\bibnamefont {Breschi}}, \bibinfo {author} {\bibfnamefont {G.}~\bibnamefont {Carullo}}, \bibinfo {author} {\bibfnamefont {S.}~\bibnamefont {Albanesi}}, \bibinfo {author} {\bibfnamefont {P.}~\bibnamefont {Rettegno}}, \bibinfo {author} {\bibfnamefont {S.}~\bibnamefont {Bernuzzi}},\ and\ \bibinfo {author} {\bibfnamefont {A.}~\bibnamefont {Nagar}},\ }\bibfield  {title} {\bibinfo {title} {Gw190521 as a dynamical capture of two nonspinning black holes},\ }\href@noop {} {\bibfield  {journal} {\bibinfo  {journal} {Nature Astronomy}\ }\textbf {\bibinfo {volume} {7}},\ \bibinfo {pages} {11} (\bibinfo {year} {2023})}\BibitemShut {NoStop}%
\end{thebibliography}%

\end{document}